\documentclass[10pt,letterpaper,twocolumn]{article} %% two column, final layout

\usepackage{ol2}
\usepackage[draft]{hyperref}
\usepackage{amsmath}
\usepackage{array}

\begin{document}

\twocolumn[ %% activate for two-column option

\title{Acousto-Optical Coherence Tomography with a digital holographic detection scheme}

%% For REVTeX it is possible to automate superscript and e-mail callouts with the superscriptaddress option; see REVTeX4 documentation.

\author{Emilie Benoit a la Guillaume,$^{1,*}$ Salma Farahi,$^1$ Emmanuel Bossy,$^1$ Michel Gross,$^2$ \\ and Francois Ramaz$^1$}

\address{
$^1$Institut Langevin, ESPCI ParisTech, CNRS UMR 7587, INSERM U979, Universit$\acute{e}$ Paris VI - Pierre et Marie Curie, \\10 rue Vauquelin 75231 Paris Cedex 05, France
\\
$^2$Laboratoire Charles Coulomb, CNRS UMR 5221, Universit$\acute{e}$ Montpellier II, 34095 Montpellier, France\\
$^*$Corresponding author: emilie.benoit@espci.fr
}

\begin{abstract}Acousto-Optical Coherence Tomography (AOCT) consists in using random phase jumps on ultrasound and light to achieve a millimeter resolution when imaging thick scattering media. We combined this technique with heterodyne off-axis digital holography. Two-dimensional images of absorbing objects embedded in scattering phantoms are obtained with a good signal to noise ratio. We study the impact of the phase modulation characteristics on the amplitude of the acousto-optic signal and on the contrast and apparent size of the absorbing inclusion.\end{abstract}


 ] %% activate for two-column option

\noindent Medical imaging is a contrast issue. Different types of waves (X-rays, ultrasound, etc.) are used depending on which organ is observed. Imaging with light, e.g. for breast cancer screening, raises the problem of detecting a millimeter-sized absorbing object in a several-centimeters thick scattering medium. When using light only, like in Diffuse Optical Tomography [1], the resolution of a breast tissues image is usually around \textit{10 mm} so that doctors can hardly detect an emerging tumor. Even if another imaging technique is used for screening, like radiography, the optical information remains precious since it supplements the knowledge of the tissues with rich physiological details [2]. The idea to use acoustic waves to achieve a millimeter resolution on optical information [3] gave rise to Ultrasound modulated Optical Tomography (UOT).

UOT is based on the acousto-optic (AO) effect which enables to tag scattered light by shifting its frequency $\omega_{L}$ of the acoustic frequency $\omega_{US}$ [4]. The detection of this weak signal at $\omega_{L}\pm\omega_{US}$ provides spatially resolved optical information by scanning the acoustic source within the medium. The size of the tagging zone is determined by the shape of the acoustic wave beam, which depends on $\omega_{US}$, on the aperture and on the focal length of the transducer. A focusing ultrasound transducer typically yields \textit{1-2 mm} lateral resolution in the focal plane when used at several \textit{MHz}. However, localization of the AO effect is about 10 times less accurate along the propagation direction of the ultrasound. As a consequence, the optical and acoustic waves have to be temporally reshaped in order to improve the axial resolution.\

\begin{figure}[htb]
\centering
\includegraphics[width=7.5cm]{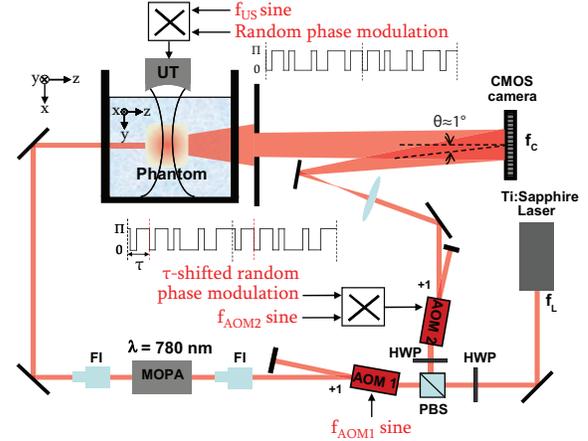}
\caption{\footnotesize{(Color on-line) Experimental set-up. HWP: half-wave plate, PBS: polarizing beam splitter, AOM1, AOM2: acousto-optic modulators, FI: Faraday isolator, UT: ultrasound transducer, $f_C$: camera framerate, $f_L$: laser frequency. In the tank, the scheme axes are modified in order to give a view of the inside. }}
\label{fig:1} 
\end{figure}

Wang and Ku [5] proposed to encode each axial position with a frequency-swept acoustic wave but the recording time was greater than the speckle decorrelation time, excluding any \textit{in vivo} experiment. As commonly used in standard echography, acoustic bursts are a solution to get axial resolution that involves times compatible with speckle decorrelation and medical standards on ultrasound exposure. The burst technique has benefited photo-refractive UOT [6,7] and has also been implemented with digital holographic detection by Atlan et al. [8]. Nevertheless, detecting µs signals with a camera whose minimum acquisition time is in the order of hundreds of $\mu s$ is quite inefficient. Lesaffre et al. solution [9], known as AOCT, is interesting because it returns to continuous acoustic and optical waves, performing a good resolution though, thanks to a random phase modulation on ultrasound and light. Up to now, AOCT technique has only been implemented in a photo-refractive holographic scheme. In this Letter, we demonstrate the potential of AOCT when combined with digital holography. We first experimentally study the influence of the random pattern characteristics on 1D AO images. Then the resolution of the technique is tested by performing 2D images on thick tissue-like phantoms containing small absorbing objects. We finally compare two AO profiles of the same object respectively obtained with AOCT and the burst technique.

Figure 1 depicts schematically the experimental set-up. A single-frequency 300-mW CW Ti:Sapphire laser (Coherent, MBR 110) with a coherence length of 300 m is tuned at 780 nm in order to work in the optical therapeutic window. A polarising beam splitter (PBS) divides light into an object beam and a reference beam. The object beam, after being amplified by a 2.5 W semiconductor amplifier (MOPA, Sacher Lasertechnik) illuminates a scattering sample immersed in a transparent water tank and insonified by a single element acoustic transducer (Panametrics A395S, $f_{US}$=2.3 MHz, focal length=78 mm, diameter=38 mm). The transmitted light passes through a 6-mm diameter circular diaphragm and interferes with the reference beam on a 12-bit, 1024x1024 pixels fast CMOS camera (Photron FastCam SA4) which records images at frame rate $f_C$ (3.6 kHz maximum frame rate at full resolution). Two AO modulators (AOM 1, 2) are placed in each arm of the interferometric set-up and shift the frequency of light of $f_{AOM1,2}$. By choosing $f_{AOM2} - f_{AOM1} = f_{US} + f_C/2$, we exclusively select the tagged photons because the other contributions vary too fast to be caught by the camera [10]  and we perform a two-phases demodulation in order to reduce the mean intensity term [11]. The off-axis configuration adds a spatial separation of modulated and non-modulated light in the Fourier domain [12].

To get axial resolution in AOCT, the same random phase modulation $\Phi(t)$ is applied to the optical reference wave and to the acoustic wave via two synchronized waveform generators. 
The reference beam is delayed of a time $\tau$ compared to the acoustic signal in order to select a probed zone localized at a given $y_0$ position within the sample ($y_0=v_{US}\tau$). The interference cross term detected by the camera is the same as in standard AO imaging apart from the phase modulation factor. It can be written as:   $\underline{E_T} \ \underline{E_R}^{*}\  \label{Eq2} \nonumber  
exp[j\phi(t-y/v_{US})] exp[-j\phi(t-\tau)] + c.c.$, where $\underline{E_T}$ and $\underline{E_R}$ are the complex amplitudes of the tagged photons field and the reference field, $v_{US}$ is the velocity of ultrasound (in water $ v_{US}\simeq 1500 \ m.s^{-1} $), and $c.c.$ is the complex conjugate. Lesaffre et al. [13] did the theoretical study of the tagged photons field in AOCT in the case of a  $\lbrace 0;\pi \rbrace$ random phase jump modulation. They demonstrated that the tagged photons field is proportional to the correlation product between the two random phase modulations, both in amplitude and spatial extent.

In our configuration of holographic detection, the same phase modulation, based on $\lbrace 0;\pi \rbrace$ phase jumps occurring every $\delta t$, is implemented. The random pattern is recorded in the memory of a waveform generator (Tektronix AFG 3252) which limits the signal length to 524 $\mu s$. As a consequence, the pattern has to be repeated several times when the camera acquisition time exceeds 524 $\mu s$. The first experiment aims at studying the dependence of the AO signal on the modulation characteristic time $\delta t$. The results are compiled in figures 2 and 3.

\begin{figure}[htb]
\centering
\includegraphics[width=7cm]{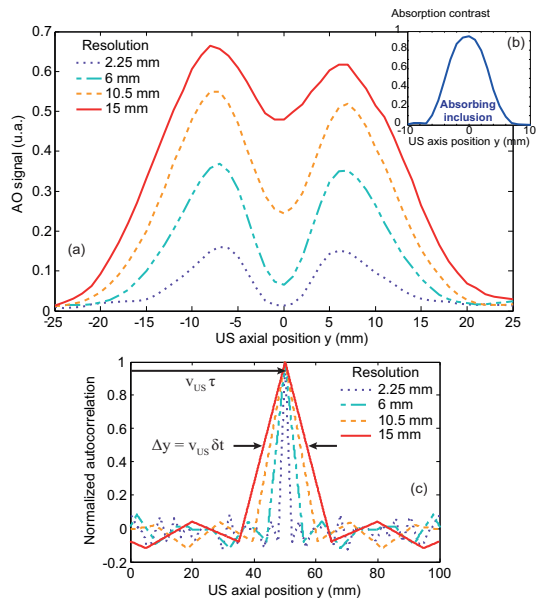}
\caption{\footnotesize{(Color online) (a) AO profiles along y axis of a scattering phantom (thickness=3 cm, $\mu_s'=10 \ cm^{-1}$), containing a black inked cylinder of  $10 \times 6 \times 6 \ mm^3 \ (x,y,z)$. The same AO profile is represented with different resolution values. $f_C$=500 Hz. (b) Shape of the absorbing inclusion, extracted from the AO profile with a 2.25 mm resolution. (c) Numerical simulation of the normalized autocorrelation function of the phase modulation sequences used to obtain the AO profiles in (a).}}
\label{fig:2} 
\end{figure}

Figure 2(a) shows axial AO profiles obtained in a 3-cm thick (along z axis) 10$\%$-Intralipid and agar gel containing an absorbing cylinder for several resolution values. The transport mean free path $l^*$ of this phantom is about 1 mm. The phase modulation pattern has an initial length of $200 \ \mu s$ (100 jumps, $\delta t =2 \ \mu s $) and is stretched or compressed by changing the reading period of the waveform generator in order to monitor the characteristic time and thus the axial resolution. The typical shape of the autocorrelation function of the random phase jump sequence is a triangle of $\Delta y=v_{US} \delta t$ in width at half maximum, as shown in figure 2(c). The theoretical minimum for $\Delta y$ is one period of the acoustic wave, i.e. 0.65 mm at $f_{US}$=2.3 MHz.  When $\delta t$ increases from 1.5 to 10 $\mu s$, the peak width of the autocorrelation function becomes larger so that the absorbing inclusion appears in the AO profile with a contrast decreasing from 95 to 54$\%$. Figure 3 shows the evolution of the amplitude of the AO signal and the inclusion contrast and size as a function of the resolution. These curves are extracted from the experimental measurements. In order to calculate the characteristics of the inclusion revealed by the AO profiles (contrast and size), each profile is fitted with a Gaussian envelope used as an approximation of the scattering photons intensity profile along the acoustic axis. The difference between the envelope and the measured profile reveals the inclusion shape as seen in figure 2(b).

\begin{figure}[htb]
\centering
\includegraphics[width=8cm]{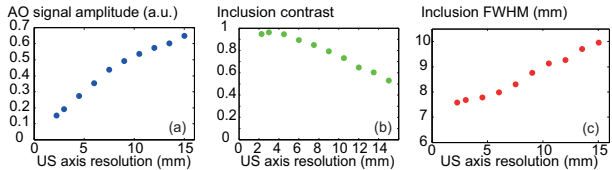}
\caption{\footnotesize{Characteristics of the AO profile as functions of the resolution: (a) amplitude, (b) contrast, (c) full width at half maximum of the inclusion (FWHM). }}
\label{fig:3} 
\end{figure}

Figure 3 demonstrates that the AO signal amplitude and the contrast of the inclusion vary in opposite directions as functions of $\Delta y$, which is expected since the peak of the autocorrelation function of the random phase pattern grows in amplitude and widens when $\Delta y$ increases. Thus, the difficulty to perform high contrast AO profiles lies in dealing with weak signals. As long as the resolution is less than the absorbing object size, the contrast and the FWHM of the inclusion are nearly constant and give an estimation of the inclusion diameter. If we look at the first points on graph 3(c), the extracted diameter of the cylinder is 7.5 mm which is 25$\%$ as large as the real one (6 mm). This difference can be explained by the inhomogeneity of the photons distribution in the presence of an absorbing object. Indeed, the probability of photon absorption is larger in the neighbourhood of the inclusion than in the rest of the scattering medium.

\begin{figure}[htb]
\centering
\includegraphics[width=8cm]{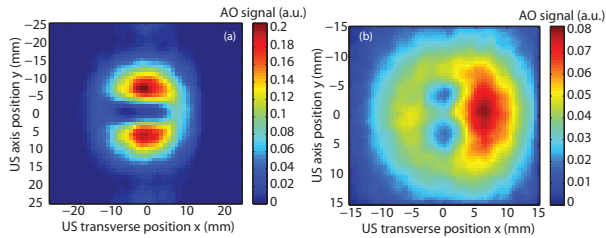}
\caption{\footnotesize{(Color online) (a) AO image of the same phantom as in figure 2, performed with $\Delta y=3 \ mm$ axial resolution, $f_C$=500 Hz. (b) AO image of a xy section of a scattering phantom (thickness=4 cm, $\mu_s'=10 \ cm^{-1}$) with 2 absorbing inclusions of $3\times 3\times 5 \ mm^3 \ (x,y,z)$, performed with $\Delta y=2.25 \ mm$ axial resolution, $f_C$=500 Hz.}}
\label{fig:4} 
\end{figure}

An xy AO image of the same sample, presented in figure 4(a), is obtained by scanning both the US transducer along x and the delay $\tau$ between the two random phase patterns. The quality of this imaging technique has also been tested by performing a 2D AO image of a 4-cm thick 10$\%$-Intralipid and agar scattering phantom ($l^*=1 \ mm$) containing two black-inked absorbing inclusions separated by 3.5 mm (figure 4(b)). The random pattern jump time is $\delta t= 1.5 \ \mu s$, giving a spatial resolution of 2.25 mm. The two inclusions are clearly separated and their size is well retrieved.
 
\begin{figure}[htb]
\centering
\includegraphics[width=6.5cm]{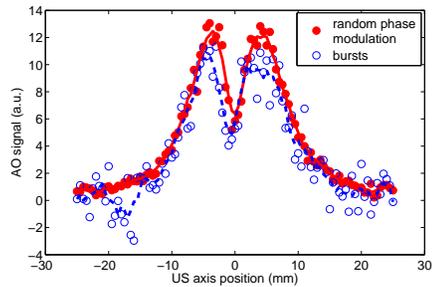}
\caption{\footnotesize{AO profile along the acoustic axis of a 10$\%$-Intralipid and agar scattering phantom (thickness=2 cm, $\mu_s'=10 \ cm^{-1}$) with a black-inked absorbing inclusion of $3 \times 3 \times 6 \ mm^{3} \ (x,y,z)$, performed with $\Delta y=2.6 \ mm$ axial resolution thanks to random phase jumps (filled dots) or bursts (empty dots), $f_C$=1 kHz.}}
\label{fig:5} 
\end{figure}
 
The advantage of using random phase jumps instead of bursts to achieve axial resolution is the noise reduction on the AO signal. In figure 5, the same scattering phantom is scanned along the acoustic axis with both techniques in the same conditions of acoustic tagging. As a consequence, the signal amplitude, the inclusion contrast and the FWHM are equal within less than 10$\%$ on the two profiles. However, the signal-to-noise ratio on the AO profile is 3 times worse with the burst technique because the useful signal represents only a short fraction of the acquisition time and is thus surrounded by a parasitic signal.

In conclusion, we have successfully implemented a random phase modulation on US and light to perform digital holographic UOT with a millimeter resolution. The experiments have been run with continuous acoustic and optical waves but the next step is to apply  AOCT with a 1-ms-pulse laser in order to keep a high level of signal while observing medical safety standards. 

The authors gratefully acknowledge the \textit{Agence Nationale de la Recherche} for financial support (project ICLM-ANR-2011-BS04-017-01).

%\pagebreak

\pagebreak

\section*{Informational Fourth Page}

\end{document}